\documentstyle[12pt,epsfig]{article}
\newcommand{\lsim}{\raisebox{-0.13cm}{~\shortstack{$<$ \\[-0.07cm] $\sim$}}~}
\newcommand{\gsim}{\raisebox{-0.13cm}{~\shortstack{$>$ \\[-0.07cm] $\sim$}}~}

\begin{document}

\begin{flushright}
THES-TP 2001/03 \\
 PM/01-15
\end{flushright}

\vspace{1cm}

\begin{center}

{\large\sc {\bf Charge and Color Breaking conditions associated
with the top quark Yukawa coupling }}

\vspace{1cm} {C. Le Mou\"el~\footnote{Electronic address:
lemouel@physics.auth.gr}} \vspace{.5cm}

 {Dept. of Theoretical Physics, Aristotle University of
Thessaloniki,\\ 54006 Thessaloniki, Greece.}

\vspace{.5cm}

{Physique Math\'ematique et Th\'eorique, UMR No 5825--CNRS, \\
Universit\'e Montpellier II, F--34095 Montpellier Cedex 5,
France.}

\end{center}

\vspace*{1.5cm}

\begin{abstract}
\setlength{\baselineskip}{15pt}

In the Minimal Supersymmetric extension of the Standard Model
(MSSM), the Charge and Color Breaking (CCB) vacuum in the plane
$(H_2,\tilde{t}_L, \tilde{t}_R)$ typically deviates largely from
all D-flat directions, due to the large effects induced by the
presence of the top quark Yukawa coupling. As a result, the
critical CCB bound on the trilinear soft term $A_t$ becomes more
restrictive than the D-flat bound $A_t^2 \le 3
(m_{\tilde{t}_L}^2+m_{\tilde{t}_R}^2+m_2^2)$. For large $\tan
\beta$, we consider the effect of a splitting between the soft
squark masses $m_{\tilde{t}_L},m_{\tilde{t}_R}$ on this optimal
CCB bound and give a useful approximation for it, accurate within
$1 \%$ in all interesting phenomenological cases. The physical
implications on the top squark mass spectrum and the one-loop
upper bound on the lightest CP-even Higgs boson mass are also
discussed in a model-independent way.

\end{abstract}

\newpage
\setlength{\baselineskip}{15pt}
 \setcounter{footnote}{0}

\section{Introduction}

In the Minimal Supersymmetric Standard Model (MSSM), spontaneous
symmetry breaking may occur into a dangerous Charge and Color
Breaking (CCB) vacuum, therefore, destabilizing the realistic
physical ElectroWeak (EW) vacuum \cite{ccbf,ccbcas}. To avoid this
danger, CCB conditions must be imposed on the soft supersymmetry
(SUSY) breaking terms which enter the scalar potential of the
MSSM. Such conditions and their physical consequences have been
extensively discussed in the literature in a large variety of
directions in the MSSM scalar field space [see, e.g.,
\cite{ccbf,ccbcas,ccb1} and references therein]. CCB vacua
associated
 with the top quark Yukawa coupling represent a special
class of dangerous vacua in the sense that they typically deviate
largely from D-flat directions and can furthermore develop in the
close vicinity of the EW vacuum. Such features were first
investigated in ref.\cite{ccbcas} in the field direction
$(H_1,H_2,\tilde{t}_L, \tilde{t}_R, \tilde{\nu}_L)$, neglecting
possible deviations from the $SU(3)_c$ D-flat direction. In
\cite{ccb1}, we considered the restricted plane $(H_2,\tilde{t}_L,
\tilde{t}_R)$ and proposed a new method to evaluate analytically,
in a fully model-independent way, the Vacuum Expectation Values
(VEVs) of the CCB vacuum \cite{ccb1}. We showed that the effect of
deviations from the $SU(3)_c$ D-flat direction, typically small in
minimal supergravity (mSUGRA) models [as considered in
ref.\cite{ccbcas}], cannot be neglected in a model-independent way
and tend to make more restrictive the CCB condition on the
trilinear soft term $A_t$. This new feature appears in models with
a large splitting at the EW scale between the soft squark masses
$m_{\tilde{t}_L},m_{\tilde{t}_R}$, as occurs for instance in some
string effective field theories \cite{ans,anom}. For large $\tan
\beta$ and $m_{A^0} \gg m_{Z^0}$, it was also pointed out that the
one-loop Higgs maximal mixing for the top squark masses is largely
ruled out by CCB considerations \cite{ccb1}, assuming a common
soft squark mass $M_{SUSY}=m_{\tilde{t}_L}=m_{\tilde{t}_R}$, and
taking furthermore a simplified value for the Higgs maximal mixing
 $A_t=\sqrt{6} \ m_{\tilde{t}}$, which is accurate only for $m_{\tilde{t}}
\equiv \sqrt{M_{SUSY}^2+m_t^2} \gg m_t$ \cite{higgs1,higgs2}.\\
 In the present paper, we first
summarize our method to evaluate the optimal CCB condition on
$A_t$ in the plane $(H_2,\tilde{t}_L, \tilde{t}_R)$ and extend
then the latter result by incorporating the possibility of a large
mass splitting between the soft squark masses
$m_{\tilde{t}_L},m_{\tilde{t}_R}$, taking also for comparison the
exact one-loop Higgs maximal mixing. We still concentrate on the
asymptotic regime $\tan \beta=+ \infty$, which actually provides a
benchmark CCB bound on $A_t$ with properties useful for
phenomenological applications. In particular, in the extended
plane $(H_1, H_2,\tilde{t}_L, \tilde{t}_R)$, this benchmark value
proves to put an upper bound on the CCB allowed values for the top
squark mixing term $|\tilde{A}_t| \equiv |A_t+\mu/\tan \beta|$
\cite{ccb3}. We present an analytic approximation for it, accurate
within $1 \ \%$ in all interesting phenomenological cases.
Finally, we consider some physical implications of this bound,
including the effect on the top squark mass spectrum and the
one-loop upper bound on the lightest CP-even Higgs boson mass.

\section{The CCB vacuum in the plane $(H_2,\tilde{t}_L, \tilde{t}_R)$}

At the tree-level, the effective potential in the plane
$(H_2,\tilde{t}_L, \tilde{t}_R)$ reads \cite{ccbf,ccbcas,ccb1}:
\begin{eqnarray}
\label{V3ch} V_3&=&m_2^2 H_2^2+m_{\tilde{t}_L}^2 {\tilde{t}_L}^2+
m_{\tilde{t}_R}^2 {\tilde{t}_R}^2
 -2 Y_{t} A_t H_2 \tilde{t}_L \tilde{t}_R
+Y_t^2 (H_2^2 {\tilde{t}_L}^2+H_2^2 {\tilde{t}_R}^2+
{\tilde{t}_L}^2 {\tilde{t}_R}^2) \nonumber \\ &&
+{\frac{g_1^2}{8}} (H_2^2+{\frac{{\tilde{t}_L}^2}{3}}- {\frac{4
{\tilde{t}_R}^2}{3}})^2+{\frac{g_2^2}{8}}
(H_2^2-{\tilde{t}_L}^2)^2+ {\frac{g_3^2}{6}}
({\tilde{t}_L}^2-{\tilde{t}_R}^2)^2
\end{eqnarray}
where $H_2$ denotes the neutral component of the corresponding
Higgs scalar $SU(2)_L$ doublet, and $\tilde{t}_L$, $\tilde{t}_R$
are, respectively, the left and right top squark fields. All
fields are supposed to be real. $H_2,\tilde{t}_L$, the top quark
Yukawa coupling $Y_t$ and the trilinear soft term $A_t$ are also
assumed to be positive, which can be arranged by a phase
redefinition of the fields. Finally, positivity for the squared
soft squark masses $m_{\tilde{t}_L}^2,m_{\tilde{t}_R}^2$ is
assumed to avoid an obvious instability of the potential at the
origin of the fields.\\
 As is well-known, this potential becomes negative in the
D-flat direction $|H_2|=|\tilde{t}_L|=|\tilde{t}_R|$, unless the
condition \cite{ccbf,ccbcas}:
\begin{equation}
\label{condfrere} A_t^2 \le (A_t^D)^2 \equiv 3
(m_{\tilde{t}_L}^2+m_{\tilde{t}_R}^2+m_2^2)
\end{equation}
is verified. If the top quark Yukawa coupling were as small as the
Yukawa couplings of the first two generations of quarks, this
relation would provide an accurate necessary and sufficient
condition to avoid CCB in the plane $(H_2,\tilde{t}_L,
\tilde{t}_R)$ \cite{ccbf,ccbcas,ccb1}. But this is not the case
and large deviations of the CCB vacuum from all D-flat directions
are typically observed, making more restrictive the critical CCB
bound on $A_t$. In fact, looking at the extremal equations
associated with the potential $V_3$, eq.(\ref{V3ch}), one finds
that the minimum of the potential lies in the D-flat direction
only for:
\begin{equation}
\label{massrel} m_{\tilde{t}_L}^2=m_{\tilde{t}_R}^2=m_2^2 \ .
\end{equation}
Violations of these relations trigger deviations from the D-flat
directions \cite{ccb1}. In particular, the first relation in
eq.(\ref{massrel}) is intimately related to the deviation from the
$SU(3)_c$ D-flat direction \cite{ccb1}, previously disregarded in
\cite{ccbcas}. We can conveniently keep track of this feature by
defining $f \equiv \tilde{t}_R/\tilde{t}_L$. Alignment of the CCB
vacuum in the $SU(3)_c$ D-flat direction will correspond to
$<f>=\pm 1$.\\
  Replacing $\tilde{t}_R
\rightarrow f \ \tilde{t}_L$ in $V_3$, eq.(\ref{V3ch}), which is
unambiguous provided by $\tilde{t}_L \neq 0$, the extremal
equation associated with a non-trivial CCB VEV $<\tilde{t}_L>$ is
straightforwardly solved, and gives \cite{ccb1}:
\begin{equation}
\label{eqUL} <\tilde{t}_L>^2= -2 \frac{B_3}{A_3}
\end{equation}
with
\begin{eqnarray}
 B_3&=&H_2^2 \frac{(12 Y_t^2-4 g_1^2) f^2+12 Y_t^2+ g_1^2-3 g_2^2}{12} -2 A_t Y_t f
H_2+ m_{\tilde{t}_L}^2+f^2 m_{\tilde{t}_R}^2 \\
 A_3 &=& 4
Y_t^2 f^2 +\frac{g_1^2(4 f^2-1)^2}{18}+\frac{g_2^2 f^4}{2}+\frac{2
g_3^2 (f^2-1)^2}{3}
\end{eqnarray}
where the field parameters $H_2,f$ take their vacuum expectation
values $H_2=<H_2>$, $f=<f>$. Consistency of this solution requires
that $B_3 \le 0$, implying on one hand $<f> \ \ge 0$ and therefore
$<H_2>,<\tilde{t}_{L/R}> \ \ge 0$ \cite{ccb1}. On the other, it is
easily shown that for $A_t \le A_t^{(0)}$, where
\begin{equation}
\label{condsuf} A_t^{(0)} \equiv m_{\tilde{t}_L}
\sqrt{1-{\frac{g_1^2}{3 Y_t^2}}}+m_{\tilde{t}_R}
\sqrt{{1-{\frac{(3 g_2^2-g_1^2)}{12 Y_t^2}}}} \simeq
m_{\tilde{t}_L}+m_{\tilde{t}_R}
\end{equation}
the global minimum of the potential $V_3$, eq.(\ref{V3ch}), is
automatically trapped in the plane $\tilde{t}_R=\tilde{t}_L=0$ and
cannot be lower than the EW vacuum [In this situation, we have
always $B_3 \ge 0$, whatever $<f>$ is]. The bound $A_t^{(0)}$
therefore provides a very simple sufficient condition on $A_t$ to
avoid CCB in the plane $(H_2, \tilde{t}_L, \tilde{t}_R)$
\cite{ccb1}. It is also restrictive enough to secure some
interesting model-dependent scenarios. This includes the infrared
quasi-fixed point scenario in an mSUGRA context, for both low and
large $\tan \beta$ \cite{ccb1}. In a quite different context, we
also mention the effective MSSM recently proposed, coming from an
underlying model where the top/stop sector is living in the bulk
of an extra dimension \cite{delga}. In this case, the trilinear
soft term $|A_t|$ is related to the soft squark mass
$m_{\tilde{t}_R}$, with $|A_t|=m_{\tilde{t}_R}$. A quick look at
eq.(\ref{condsuf}) shows that this model is also exempt from a CCB
vacuum in the plane $(H_2, \tilde{t}_L, \tilde{t}_R)$.\\

The previous sufficient bound to avoid CCB in the plane $(H_2,
\tilde{t}_L, \tilde{t}_R)$ can be improved. This requires some
information on the remaining extremal equations associated with
$H_2$ and $f$. Taking $\tilde{t}_L=<\tilde{t}_L>$ in the potential
$V_3$, eq.(\ref{V3ch}), these equations are straightforwardly
obtained. The derivative with respect to $f$ provides an equation
quadratic in $H_2$ and quartic in $f$
\begin{equation}
\label{eqf3} a_3 f H_2^2+ b_3 H_2+c_3 f=0
\end{equation}
where $a_3=36 Y_t^4 (-1+f^2)+O(g_i^2), b_3=O(g_i^2), c_3=-36 Y_t^2
(m_{\tilde{t}_L}^2-f^2 m_{\tilde{t}_R}^2)+O(g_i^2)$ [the exact
value of the coefficients $a_3,b_3,c_3$ can be found in
\cite{ccb1}]. Numerical investigation shows that, at the CCB
vacuum, the gauge contributions $\sim O(g_i^2)$ to the
coefficients $a_3,b_3,c_3$ can be safely neglected. Doing so, the
extremal equation, eq.(\ref{eqf3}), is solved exactly and gives
\begin{equation}
\label{approxf}
 <f>\simeq \sqrt{{\frac{m_{\tilde{t}_L}^2+ Y_t^2
<H_{2}>^2}{m_{\tilde{t}_R}^2+Y_t^2 <H_{2}>^2}}}
\end{equation}
A more transparent approximation of the deviation parameter $<f>$,
independent of the CCB VEV $<H_2>$, can be subsequently derived
\cite{ccb1}:
\begin{equation}
\label{f0ap3} <f> \simeq f^{(0)}_3 \equiv \sqrt{{\frac{A_t^2+2
m_{\tilde{t}_L}^2-m_{\tilde{t}_R}^2}{A_t^2+2
m_{\tilde{t}_R}^2-m_{\tilde{t}_L}^2}}}
\end{equation}
The accuracy of this approximated value proves to be excellent,
whatever the values of the soft parameters
$A_t,m_{\tilde{t}_L},m_{\tilde{t}_R}$ are: it fits the exact
result $<f>$ within $5 \%$, or even less for not a too large
splitting between the soft squark masses
$m_{\tilde{t}_L},m_{\tilde{t}_R}$ \cite{ccb1}. It also clearly
indicates that alignment of the CCB vacuum in the $SU(3)_c$ D-flat
direction is a model-dependent feature which occurs in two
different cases:\\
 - For $A_t \gg 2
Max[m_{\tilde{t}_L},m_{\tilde{t}_R}]$. However, $A_t$ is also very
large compared to the critical CCB bound on $A_t$ above which CCB
occurs.\\
 - For $m_{\tilde{t}_L}^2=m_{\tilde{t}_R}^2$ [the first
 mass relation in eq.(\ref{massrel})]. Then, the potential $V_3$, eq.(\ref{V3ch}),
 has an underlying approximate symmetry $\tilde{t}_L \leftrightarrow \tilde{t}_R$
 broken by tiny $O(g_1^2,g_2^2)$ contributions. Therefore, any non-trivial CCB
extremum is necessarily nearly aligned in the $SU(3)_c$ D-flat
direction.\\
 Model-building may favor the latter situation. This occurs
approximately, e.g., in mSUGRA models \cite{sugra}. In
ref.\cite{ccbcas}, such models were investigated, neglecting in a
first approximation the effect of the deviation from the $SU(3)_c$
D-flat direction and CCB conditions [evaluated in the extended
plane $(H_1, H_2, \tilde{t}_L, \tilde{t}_R, \tilde{\nu}_L)$] were
given in terms of the universal soft SUSY breaking parameters at
the GUT scale. However, in a model-independent way, the
possibility of a large splitting between the soft squark masses is
not ruled out and is even favored in some interesting alternatives
to the mSUGRA scenarios. Some of them incorporate a substantial
amount of non-universality between the soft SUSY breaking terms
\cite{ans,anom,sst,anom1} and possibly show large violations of
the relation $m_{\tilde{t}_L}=m_{\tilde{t}_R}$ at the EW scale.
This occurs, e.g., in some anomaly mediated scenarios
\cite{ans,anom}. In this case, the effect of the deviation from
the $SU(3)_c$ D-flat direction on the critical CCB bound cannot be
neglected, as will be illustrated in the following, and must be
properly taken into account.\\

The extremal equation associated with $H_2$ is cubic in $H_2$ and
quartic in $f$. It reads \cite{ccb1}:
\begin{equation}
\label{eqH23} \alpha_3 H_2^3+ \beta_3 H_2^2+\gamma_3
H_2+\delta_3=0
\end{equation}
where
\begin{eqnarray}
\label{alpha}
 \alpha_3 &=&-36 Y_t^4 (f^2+1)^2+[3 g_3^2
(g_1^2+g_2^2)+4 g_1^2 g_2^2] (f^2-1)^2 \nonumber \\ &&+ 6 Y_t^2
g_1^2 (4 f^4+6 f^2-1)+18 Y_t^2 g_2^2 (2 f^2+1)  \\ \label{beta}
 \beta_3&=&9 A_t Y_t f [(12 Y_t^2-4 g_1^2) f^2+12 Y_t^2+g_1^2-3 g_2^2]  \\
 \label{gamma}
  \gamma_3&=&-72 A_t^2 f^2 Y_t^2-3
(m_{\tilde{t}_L}^2+f^2 m_{\tilde{t}_R}^2) [(12 Y_t^2-4 g_1^2)
f^2+12 Y_t^2+g_1^2-3 g_2^2] \nonumber
\\ && + m_2^2 [72 Y_t^2 f^2+g_1^2 (4 f^2-1)^2+9 g_2^2+12 g_3^2
(f^2-1)^2] \\
 \label{delta}
\delta_3 &=&36 A_t Y_t f (m_{\tilde{t}_L}^2+f^2 m_{\tilde{t}_R}^2)
\end{eqnarray}
This equation, considered as a cubic polynomial in $H_2$, may be
solved exactly. In a compact trigonomical form, the CCB VEV
$<H_2>$ reads:
\begin{eqnarray}
\label{vH2}
 &&<H_2>=-{\frac{\beta_3}{3 \alpha_3}} (1-2
\frac{\sqrt{-{\cal{N}}}}{\beta_3} cos[\frac{\phi+4 \pi}{3}]) \\ &&
\ \ \
 cos \ \phi=-\frac{{\cal{M}}}{2 \sqrt{-{\cal{N}}^3}}\ \ \  ,  \ \
 \
\phi \in [0,\pi]
\end{eqnarray}
where
\begin{eqnarray}
\label{coef}
 {\cal{M}} & \equiv & 2 \beta_3^3-9
\alpha_3 \beta_3 \gamma_3+ 27 \alpha_3^2 \delta_3 \\ \label{coef1}
 {\cal{N}} &=& -\beta_3^2+3 \alpha_3 \gamma_3
 \end{eqnarray}
where the coefficient $\alpha_3,\beta_3,\gamma_3,\delta_3$,
eqs.(\ref{alpha}-\ref{delta}), should be evaluated with $f=<f>$.
For clarity, let us add some comments concerning this solution:\\
 - It can be shown that a CCB vacuum may develop in the plane
 $(H_2, \tilde{t}_L, \tilde{t}_R)$
 only if the extremal equation eq.(\ref{eqH23}) has three real roots in
 $H_2$. Then, these roots are found to be positive and the
 CCB VEV $<H_2>$ is given by the intermediate one \cite{ccb1}, which is written in its
trigonomical analytic form in eqs.(\ref{vH2}-{\ref{coef1}).
$<H_2>$ depends on $<f>$, for which we have an accurate
approximation $<f> \simeq f^{(0)}_3$, eq.(\ref{f0ap3}). This way,
we obtain in turn an accurate approximation for $<H_2>$, which
fits the exact result within less than $1 \%$. This accuracy is
enough in particular to evaluate the critical CCB bound on $A_t$
with a precision of order $ \sim 1 \ GeV$\footnote{This precision
can be improved at will with the iterative procedure proposed in
\cite{ccb1}.}.\\
 - The extremal
 equation, eq.(\ref{eqH23}), has three real roots in
 $H_2$ if (and only
 if) the following condition is verified:
 \begin{equation}
\label{condsuf1}
 {\cal{C}}_3 \equiv {\cal{M}}^2+4 {\cal{N}}^3 \le 0
\end{equation}
As noted above, no CCB vacuum may develop for ${\cal{C}}_3 \ge 0$.
Therefore, the quantity ${\cal{C}}_3$ provides an additional
criterion to avoid CCB in the plane $(H_2, \tilde{t}_L,
\tilde{t}_R)$, which enables us to improve the sufficient bound
$A_t^{(0)}$, eq.(\ref{condsuf}). Taking $f=f^{(0)}_3$, and giving
values to all parameters except $A_t$, the equation
${\cal{C}}_3=0$ can be solved numerically as a function of $A_t$.
Below the largest positive solution thus obtained, denoted
$A_t^{(1)}$, the relation ${\cal{C}}_3 \ge 0$ is always found.
Typically, the hierarchy $A_t^{(1)} \ge A_t^{(0)}$ holds, except
in the unphysical regime where the lightest stop mass is vanishing
$m_{\tilde{t}_1} \simeq 0$ \cite{ccb1}. Nevertheless, defining
\begin{equation}
\label{csuf} A_t^{suf} \equiv Max[A_t^{(0)},A_t^{(1)}] \ ,
\end{equation}
this quantity may be identified as the optimal sufficient bound on
$A_t$ below which no CCB vacuum may develop in the plane $(H_2,
\tilde{t}_L,\tilde{t}_R)$. For $A_t = A_t^{suf}$, a consistent
local CCB vacuum with real and positive VEVs $(<H_2>,
<\tilde{t}_L>,<f>)$ begins to develop and soon becomes global with
increasing $A_t$ \cite{ccb1}.\\

The evaluation of the critical CCB bound on $A_t$, above which CCB
occurs, requires some additional information on the EW vacuum. At
the tree-level, this vacuum is determined by the extremal
equations
\begin{eqnarray}
\label{eqextre}
 && {\frac{m_1^2-m_2^2 \tan^2
\beta}{\tan^2\beta-1}}-{\frac{m_{Z^0}^2}{2}}=0 \\ \label{eqextre1}
 && (m_1^2+m_2^2) \tan \beta- m_3^2 (1+ \tan^2 \beta)=0
\end{eqnarray}
where $\tan \beta \equiv v_2/v_1$ is the ratio of the VEVs
$v_1,v_2$ of the EW vacuum [with $v \equiv \sqrt{v_1^2+v_2^2}=174
\ GeV$]. For $\tan \beta \ge 1$, the minimal value of the EW
potential reads:
\begin{equation}
\label{VEW}
 <V>|_{EW}=-{\frac{[m_2^2-m_1^2+\sqrt{(m_1^2+m_2^2)^2-4 m_3^4}]^2}{2
(g_1^2+g_2^2)}}
\end{equation}
At a consistent EW vacuum, this quantity depends only on two free
parameters that may conveniently be chosen to be $\tan \beta$ and
the pseudo-scalar mass $m_{A^0}=\sqrt{m_1^2+m_2^2}$.
 The critical CCB bound $A_t^{CCB}$
 is found by comparing the depth of the potential at the EW vacuum
and at the CCB vacuum:
\begin{eqnarray}
\label{CCBcond}
 CCB & \Leftrightarrow& <V_3> \ \ \ \ < \ \ \ \ <V>|_{EW}
\\
\label{CCBcond1} & \Leftrightarrow & A_t
> A_{t}^{CCB}[m_{\tilde{t}_L},
m_{\tilde{t}_R};m_1,m_2,m_3,Y_t,g_1,g_2,g_3]
\end{eqnarray}
The maximal depth of the potential $V_3$, eq.(\ref{V3ch}), can be
given an analytical (though complicated) expression by taking the
excellent approximations of the CCB VEV's  given in
eqs.(\ref{eqUL},\ref{f0ap3},\ref{vH2}). This way, we incorporate
all possible deviations of the CCB vacuum from all D-flat
directions, including the $SU(3)_c$ D-flat one previously
disregarded \cite{ccbcas}. When all parameters are chosen [except
$A_t$], the critical CCB bound $A_{t}^{CCB}$ is straightforwardly
obtained by a numerical scan of the region $A_t \ge A_t^{suf}$. We
have $A_t ^{CCB} \simeq A_t^{suf}$, because the EW potential is
not very deep $<V>|_{EW} \sim -m_{Z^0}^4/(g_1^2+g_2^2)$, whereas
the depth of the CCB potential increases rapidly with $A_t$. This
simple procedure provides an excellent approximated value for the
critical CCB bound which fits the exact result $A_{t}^{CCB}$ with
an accuracy of order $1 \ GeV$.\\
 Let us add a few words concerning the impact of radiative corrections on
$A_{t}^{CCB}$. As is well-known, the results obtained with the
tree-level approximation of the potential may incorporate leading
one-loop corrections, provided all quantities are evaluated at an
appropriate field-dependent scale \cite{v1r,ccbcas}. This
numerical observation was, in fact, intensively used in the
context of CCB studies in order to use the relative simplicity of
the tree-level potential \cite{ccbcas}. For the EW potential at
the EW vacuum, the appropriate scale is the SUSY scale $Q_{SUSY}$,
with $Q_{SUSY} \sim M_{SUSY}\equiv
\sqrt{(m_{\tilde{t}_L}^2+m_{\tilde{t}_R}^2)/2}$ for $M_{SUSY} \gg
m_t$, whereas for $M_{SUSY} \lsim m_t$, $Q_{SUSY}$ should be taken
at a more significant SUSY mass. In the vicinity of $A_t \simeq
A_t^{suf}$ $[\simeq A_t^{CCB}]$, it can be shown that the scale
adapted to the CCB potential $V_3$, eq.(\ref{V3ch}), at the CCB
vacuum, is also the SUSY scale $Q_{SUSY}$. This result, which is
in agreement with \cite{ccbcas}, is due essentially to the fact
that the CCB vacuum proves to be rather close to the EW vacuum
\cite{ccb1}. Therefore, provided the tree-level comparison in
eq.({\ref{CCBcond}) is performed at the SUSY scale $Q_{SUSY}$, the
critical bound $A_{t}^{CCB}$ thus obtained should also incorporate
leading one-loop corrections.

\section{The CCB conditions on $A_t$ for large $\tan \beta$}

In this section, we investigate the effect of a mass splitting
between the soft squark masses $m_{\tilde{t}_L},m_{\tilde{t}_R}$
on the CCB bounds presented in Sec. II. We focus on the asymptotic
regime $\tan \beta =+ \infty$. This choice is motivated by the
nice properties of the critical CCB condition on $A_t$ in this
regime. Before turning to the numerical analysis, let us first
briefly comment on the latter.\\
 In the plane $(H_2,\tilde{t}_L, \tilde{t}_R)$, the critical CCB bound
$A_t^{CCB}$, eq.(\ref{CCBcond1}), proves to be decreasing for
increasing $\tan \beta$, and increasing for increasing $m_{A^0}$.
Moreover, in the limit $\tan \beta =+ \infty$, it can be shown
that the most restrictive CCB bound in the plane
$(H_2,\tilde{t}_L, \tilde{t}_R)$ is obtained, whatever the values
$m_{A^0}$ and $\tan \beta (\ge 1)$ are. Therefore, the benchmark
value $A_t^{CCB}|_{\tan \beta=+\infty}$ can be considered as an
optimal sufficient condition to avoid a dangerous CCB vacuum in
this plane:
\begin{equation}
\label{at1} A_t \le A_t^{CCB}|_{\tan \beta=+\infty} \Rightarrow No
\ CCB \ in \ the \ plane \ (H_2,\tilde{t}_L, \tilde{t}_R)
\end{equation}
We stress however that this interesting property does not prevent
a CCB situation from outside the plane $(H_2,\tilde{t}_L,
\tilde{t}_R)$. In particular, CCB may still occur in the extended
plane $(H_1,H_2,\tilde{t}_L, \tilde{t}_R)$, as a full
investigation of the optimal CCB conditions in this plane indeed
shows. This will be presented elsewhere \cite{ccb3}. To enlighten
the importance of the value $A_t^{CCB}$ for $\tan \beta=+\infty$,
we borrow from this study an interesting property of the CCB bound
on the stop mixing term $\tilde{A}_t \equiv A_t+\mu/\tan \beta$.
As is well-known, this quantity plays a central role in Higgs
phenomenology \cite{higgs1,higgs2,pheno2}. It can be shown that if
$|\tilde{A}_t|$ exceeds some critical value, which depends on
$m_{A^0}, \tan \beta,\mu$ and also
$m_{\tilde{t}_L},m_{\tilde{t}_R}$, CCB occurs in the plane
$(H_1,H_2,\tilde{t}_L, \tilde{t}_R)$. In the interesting
phenomenological region $m_{\tilde{t}_L},m_{\tilde{t}_R} \gsim
m_t$, one finds in addition that this critical value is maximal
for $\tan \beta=+ \infty$ and $\mu=0$ and that this maximal value
moreover coincides with the CCB bound $A_t^{CCB}|_{\tan
\beta=+\infty}$ obtained in the plane $(H_2,\tilde{t}_L,
\tilde{t}_R)$. To summarize, $A_t^{CCB}|_{\tan \beta=+\infty}$
also provides a CCB maximal mixing for the stop fields, above
which CCB unavoidably occurs in the plane $(H_1,H_2,\tilde{t}_L,
\tilde{t}_R)$:
\begin{equation} \label{at2}
 |\tilde{A}_t| \ge A_t^{CCB}|_{\tan \beta=+\infty} \Rightarrow
 CCB \ in \ the \ plane \ (H_1,H_2,\tilde{t}_L, \tilde{t}_R)
\end{equation}
This important property is our main motivation to study in detail
the numerical behaviour of this benchmark value. In the following,
we shall also give an accurate analytic approximation for it,
which should be quite useful for phenomenological applications and
to constrain model-building.\\

For $\tan \beta \rightarrow + \infty$, the EW vacuum is driven in
the plane $(H_2,\tilde{t}_L,\tilde{t}_R)$ and appears as an
additional minimum of the potential $V_3$, eq.(\ref{V3ch}), with
VEVs $v_2^2=2 m_{Z^0}^2/(g_1^2+g_2^2)$, $<\tilde{t}_{L/R}>=0$,
giving $<V>|_{EW}=-m_{Z^0}^4/2(g_1^2+g_2^2)$. Stability of the EW
vacuum in the plane $(\tilde{t}_L,\tilde{t}_R)$ is required,
otherwise an obvious CCB situation will occur. Therefore, we may
write a new CCB condition on $A_t$, which merely encodes the
physical requirement of avoiding a tachyonic lightest stop mass
\cite{ccb1}:
\begin{equation}
\label{Ameta1}
 A_t \le A_t^{inst} \equiv \frac{\sqrt{(6 m_{\tilde{t}_L}^2+ 6
 m_t^2+m_{Z^0}^2-4 m_{W^\pm}^2) (3 m_{\tilde{t}_R}^2+ 3
 m_t^2+2 m_{W^\pm}^2)}}{3 \sqrt{2} m_t}
\end{equation}
\begin{figure}[htb]
\vspace*{-1.8cm}
\begin{center}
\mbox{ \psfig{figure=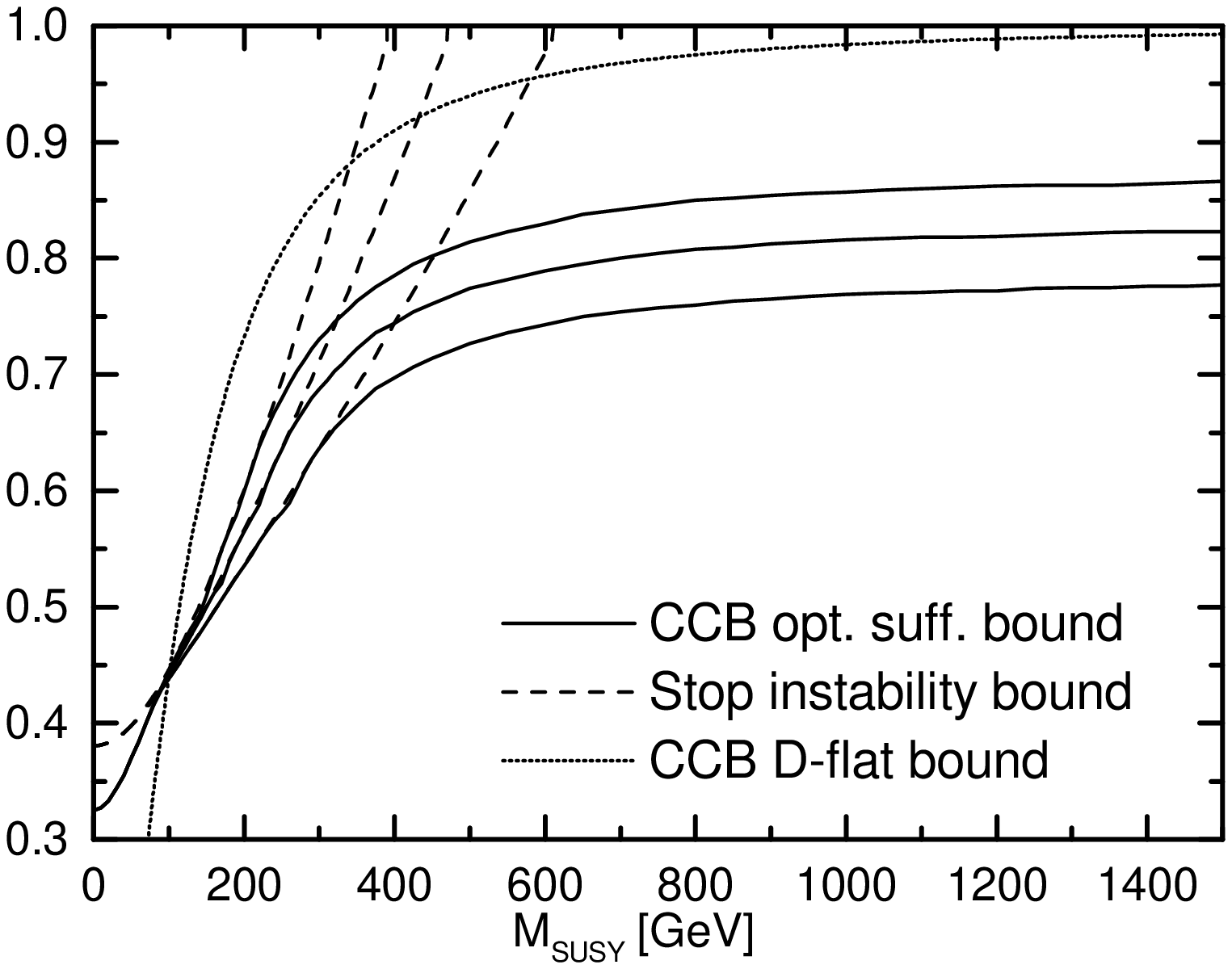,width=12.5cm}}
\end{center}
\vspace*{-3cm} \caption[Figure 4b]{The CCB optimal sufficient
bound $A_t^{suf}$, the D-flat bound $A_t^{D}$ and the instability
bound $A_t^{inst}$ versus $M_{SUSY}$. All bounds are normalized to
$\sqrt{6} m_{\tilde{t}}$. The higher, intermediate and lower lines
correspond, respectively, to $r=1,2$ and $3$.}
\end{figure}
As we will see, unlike the D-flat bound $A_t^{D}$,
eq.(\ref{condfrere}), the sufficient bound $A_t^{suf}$,
eq.(\ref{csuf}), and the CCB maximal mixing $A_t^{CCB}$,
eq.(\ref{CCBcond1}), automatically fulfill this requirement.
Notice that the instability bound $A_t^{inst}$ is only a function
of the soft squark masses $m_{\tilde{t}_L},m_{\tilde{t}_R}$, the
top mass and the gauge boson masses. This result actually extends
to the CCB bounds $A_t^{suf}$ and $A_t^{CCB}$ [for $\tan \beta
\rightarrow + \infty$]. In our numerical analysis, we take
$m_t=175 \ GeV$ and display the CCB bounds  as a function of an
average of the soft squark masses $M_{SUSY}\equiv
\sqrt{(m_{\tilde{t}_L}^2+m_{\tilde{t}_R}^2)/2}$, for three
different values of the splitting parameter $r \equiv
m_{\tilde{t}_L}/m_{\tilde{t}_R}=1,2,3$. The latter parameter will
enable to conveniently survey the effect of the CCB vacuum
deviation from the $SU(3)_c$ D-flat direction. We remark that
splitting parameters as large as $r=2$ can be found, for instance,
in anomaly mediated scenarios \cite{anom}. To be exhaustive, we
consider also the possibility of a very large splitting term
$r=3$.\\

In Fig.1, we display the optimal sufficient bound $A_t^{suf}$,
eq.(\ref{csuf}), the traditional bound in the D-flat direction
$A_t^D$, eq.(\ref{condfrere}), and the instability bound
$A_t^{inst}$, eq.(\ref{Ameta1}), as a function of $M_{SUSY}$. All
bounds are normalized to $\sqrt{6} m_{\tilde{t}}$, where
$m_{\tilde{t}} \equiv \sqrt{M_{SUSY}^2+m_t^2}$.\\
 We note first
that the sufficient bound $A_{t}^{suf}$ automatically fulfills the
important requirement of avoiding a tachyonic lightest stop. We
have always $A_t^{inst} \ge A_{t}^{suf}$. A large region [which
depends on $r$] is found where the relation $A_t^{inst}=
A_{t}^{suf}$ holds, implying that no dangerous CCB vacuum may
exist unless the EW vacuum is unstable. In this interference
regime, the EW and the CCB vacua actually overlap and the CCB VEV
$<f>$ proves to be connected quite simply to the stop mixing angle
$\tilde{\theta}$, with the relation $<f>= \tan \tilde{\theta}$
\cite{ccb1}. Here, the slightest deviation of the CCB vacuum from
the $SU(3)_c$ D-flat direction must be taken into account in the
evaluation of the CCB condition, to avoid a tachyonic stop mass.
In contrast, the D-flat bound $A_t^D$ has a rather bad behaviour.
In particular, for low $M_{SUSY}$, notice that it is not
restrictive enough to avoid a tachyonic lightest stop.\\
 Comparing the D-flat
bound $A_t^D$ with the sufficient bound $A_t^{suf}$ for $r=1,2,3$,
a precise indication of the effect of the CCB vacuum deviation
from all D-flat directions can be obtained. For large $M_{SUSY}$,
we observe first that these bounds enter an asymptotic regime,
with $A_t^D$ larger than $A_t^{suf}$. For $r=1$, the CCB vacuum is
aligned in the $SU(3)_c$ D-flat direction. Therefore, the
discrepancy between $A_t^D$ and $A_t^{suf}|_{r=1}$ is due
essentially to the deviation from the $SU(2)_c \times U(1)_Y$
D-flat direction, triggered by the large violation of the relation
$M_{SUSY}^2=m_2^2$, eq.(\ref{massrel}). For $M_{SUSY}= 1 \ TeV$,
we have, e.g., $A_t^D-A_t^{suf}|_{r=1} \simeq 365 \ GeV$. For
$r=2,3$, large deviations of the CCB vacuum from the $SU(3)_c$
D-flat direction now occur and the sufficient bound $A_t^{suf}$ is
lowered. We have, e.g., for $M_{SUSY}=1 \ TeV$,
$A_t^D-A_t^{suf}|_{r=2} \simeq 475 \ GeV$ and
$A_t^D-A_t^{suf}|_{r=3} \simeq 585 \ GeV$, and the fraction due to
the deviation from the $SU(3)_c$ D-flat direction represents $23
\%$ for $r=2$ [$37.5 \%$ for $r=3$] of the total effect. This
illustrates clearly that this additional contribution to the CCB
condition may be important and should not be neglected for a large
splitting between $m_{\tilde{t}_L},m_{\tilde{t}_R}$.\\

In Fig.2, we display now the CCB maximal mixing $A_{t}^{CCB}$,
eq.(\ref{CCBcond1}), as a function of $M_{SUSY}$, for $r=1,2,3$.
For comparison, we remark that $A_{t}^{CCB}$ follows closely the
sufficient bound $A_t^{suf}$ displayed in Fig.1, with $A_{t}^{CCB}
\ge A_t^{suf}$ for all values of $M_{SUSY}$. In the interference
regime previously mentioned, this inequality is saturated and we
furthermore have $A_t^{CCB}=A_t^{inst}$. For $M_{SUSY} \ge m_t$,
we find $A_t^{suf} \le A_{t}^{CCB} \le 1.025 \ A_t^{suf}$ for
$r=1,2,3$, the lower value being reached for $M_{SUSY} \sim m_t$
and the larger for $M_{SUSY} \sim 1.5 \ TeV$. For large $M_{SUSY}
\gsim 700 GeV$, we observe also that the CCB bound $A_{t}^{CCB}$
enters an asymptotic regime in which only tiny variations still
occur.\\
\begin{figure}[htb]
\vspace*{-1.4cm}
\begin{center}
\mbox{ \psfig{figure=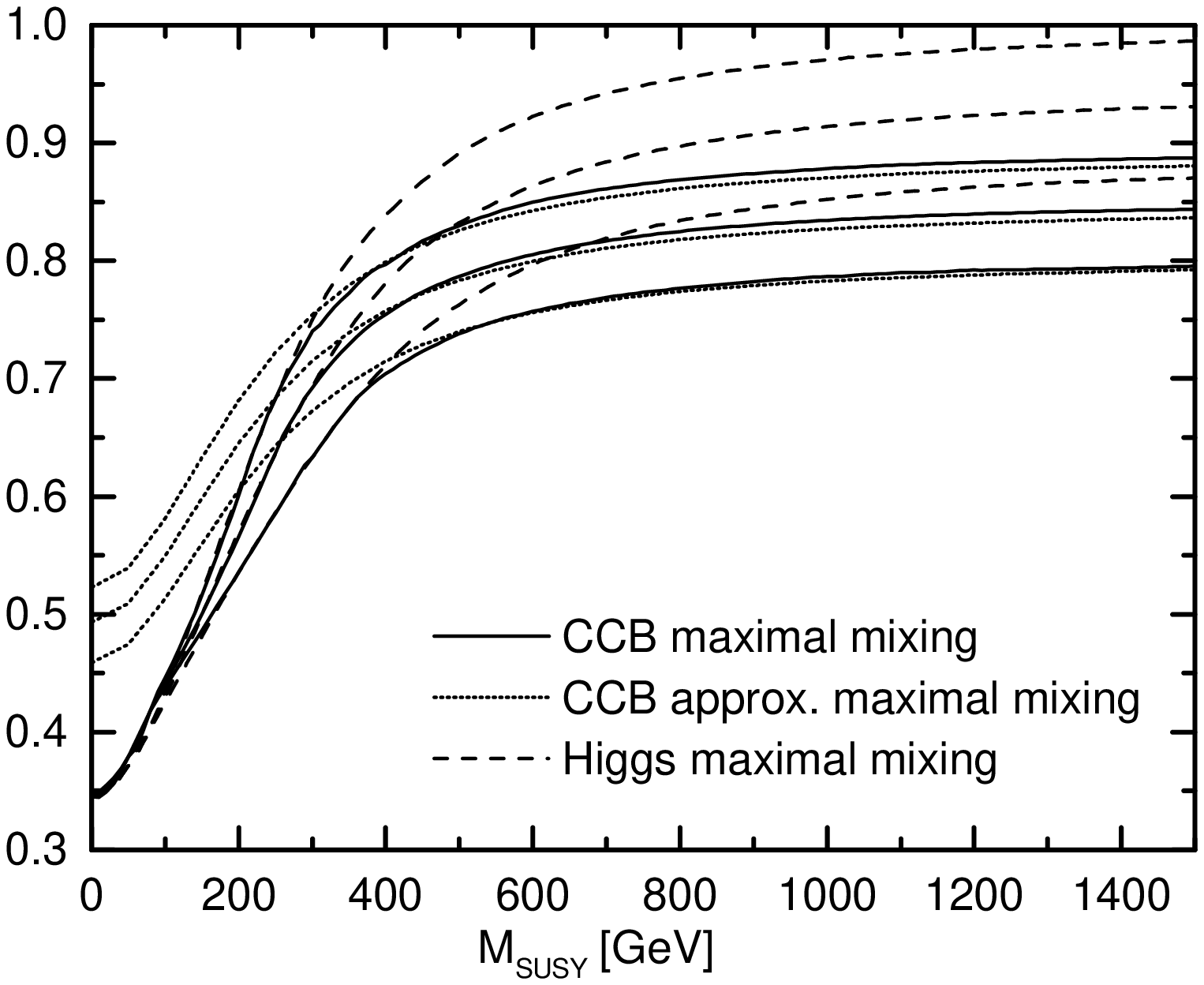,width=12.5cm}}
\end{center}
\vspace*{-1.5cm} \caption[Figure 4b]{The CCB maximal mixing
$A_t^{CCB}$, its approximation $A_t^{CCB}|_{app.}$ and the
one-loop Higgs maximal mixing $A_t^{H}$, versus $M_{SUSY}$. All
bounds are normalized to $\sqrt{6} m_{\tilde{t}}$. The higher,
intermediate and lower lines correspond, respectively, to $r=1,2$
and $3$.}
\end{figure}
In Fig.2, an approximation of the critical CCB bound $A_{t}^{CCB}$
is also displayed. It reads analytically:
\begin{equation}
\label{atap}
 A_t^{CCB}|_{app.}={\frac{2}{15}} \sqrt{{\frac{2}{3}}}
(21-r) {\frac{m_{\tilde{t}}^3}{m_{\tilde{t}}^2-{\frac{(1+r)^2
(45-42 \sqrt{3}+2 \sqrt{3} r)}{4 \sqrt{3} (21-r) r}} m_t^2}}
\end{equation}
This approximation fulfills two important requirements:
$A_t^{CCB}|_{app.} \simeq {\frac{2}{15}} \sqrt{{\frac{2}{3}}}
(21-r) m_{\tilde{t}}$ for large $m_{\tilde{t}}$, which is
consistent with the numerical behaviour observed; at first order
in $m_t/m_{\tilde{t}}$, $A_t^{CCB}|_{app.} \simeq
m_{\tilde{t}_L}+m_{\tilde{t}_R}$ for $m_{\tilde{t}_L}
m_{\tilde{t}_R} \simeq m_t^2$, as required at the center of the
interference regime [See eq.(43) in ref.\cite{ccb1}]. We stress
that this approximation holds only for $r\ge 1$. However,
numerical investigation shows that no significant variation of
$A_{t}^{CCB}$ occurs under the transformation $r \rightarrow 1/r$
\cite{ccb1}. Accordingly, for $m_{\tilde{t}_L} \le
m_{\tilde{t}_R}$, the analytical expression eq.(\ref{atap}) for
$A_t^{CCB}|_{app.}$ should be adapted by redefining $r \equiv
m_{\tilde{t}_R}/m_{\tilde{t}_L}$.\\
 For $M_{SUSY}$ large
enough, Fig.2 shows the excellent accuracy of the approximation
$A_t^{CCB}|_{app.}$. For all values of $r$, it fits the exact
result $A_t^{CCB}$ within less than $1 \ \%$. In contrast, for low
$M_{SUSY}$, it behaves rather badly. However, this feature occurs
only in a region of the parameter space where the lightest stop
mass is small, i.e. $m_{\tilde{t}_1} \lsim 100 \ GeV$ [see Fig.3].
Such a region is nearly completely excluded by experimental data
\cite{stop1}.\\
 Finally, in Fig.2 we display the one-loop Higgs maximal mixing for
 the stop masses, denoted $A_t^H$ in the following. As is well-known, the tree-level
lightest CP-even Higgs boson mass receives large one-loop
corrections from loops of top and stop quark fields, which are
essential to overcome the tree-level upper bound $m_h \le m_{Z^0}$
\cite{higgs1}. For $\tan \beta=+ \infty$ and $m_{A^0} \gg m_t$,
these corrections are maximized and we have in the top-stop
approximation \cite{higgs1}:
\begin{eqnarray}
\label{bmh}
  m_h^2&=&m_{Z^0}^2+{\frac{3 m_t^4}{8 \pi v^2}} [Log
{\frac{m_{\tilde{t}_1} m_{\tilde{t}_2}}{m_t^2}}+ \frac{4
A_t^2}{m_{\tilde{t}_2}^2-m_{\tilde{t}_1}^2} Log
{\frac{m_{\tilde{t}_2}}{m_{\tilde{t}_1}}} \nonumber \\ &&+ \frac{2
A_t^4}{(m_{\tilde{t}_2}^2-m_{\tilde{t}_1}^2)^2}
(1-\frac{m_{\tilde{t}_2}^2+m_{\tilde{t}_1}^2}{m_{\tilde{t}_2}^2-m_{\tilde{t}_1}^2}Log
{\frac{m_{\tilde{t}_2}}{m_{\tilde{t}_1}}})] \ \ \ \ , \ \ v=174 \
GeV
\end{eqnarray}
where the stop masses read:
\begin{equation}
\label{mstop}
 m_{\tilde{t}_{1},\tilde{t}_2}^2=
{\frac{m_{\tilde{t}_L}^2+m_{\tilde{t}_R}^2}{2}}+ m_t^2
-{\frac{1}{4}} m_{Z^0}^2
 \mp \sqrt{ m_t^2
A_t^2+{\frac{[6 (m_{\tilde{t}_L}^2-m_{\tilde{t}_R}^2)-8 \
m_{W^{\pm}}^2 + 5 \ m_{Z^0}^2]^2}{144}}}
\end{equation}
\begin{figure}[htb]
\vspace*{-1.4cm}
\begin{center}
\mbox{ \psfig{figure=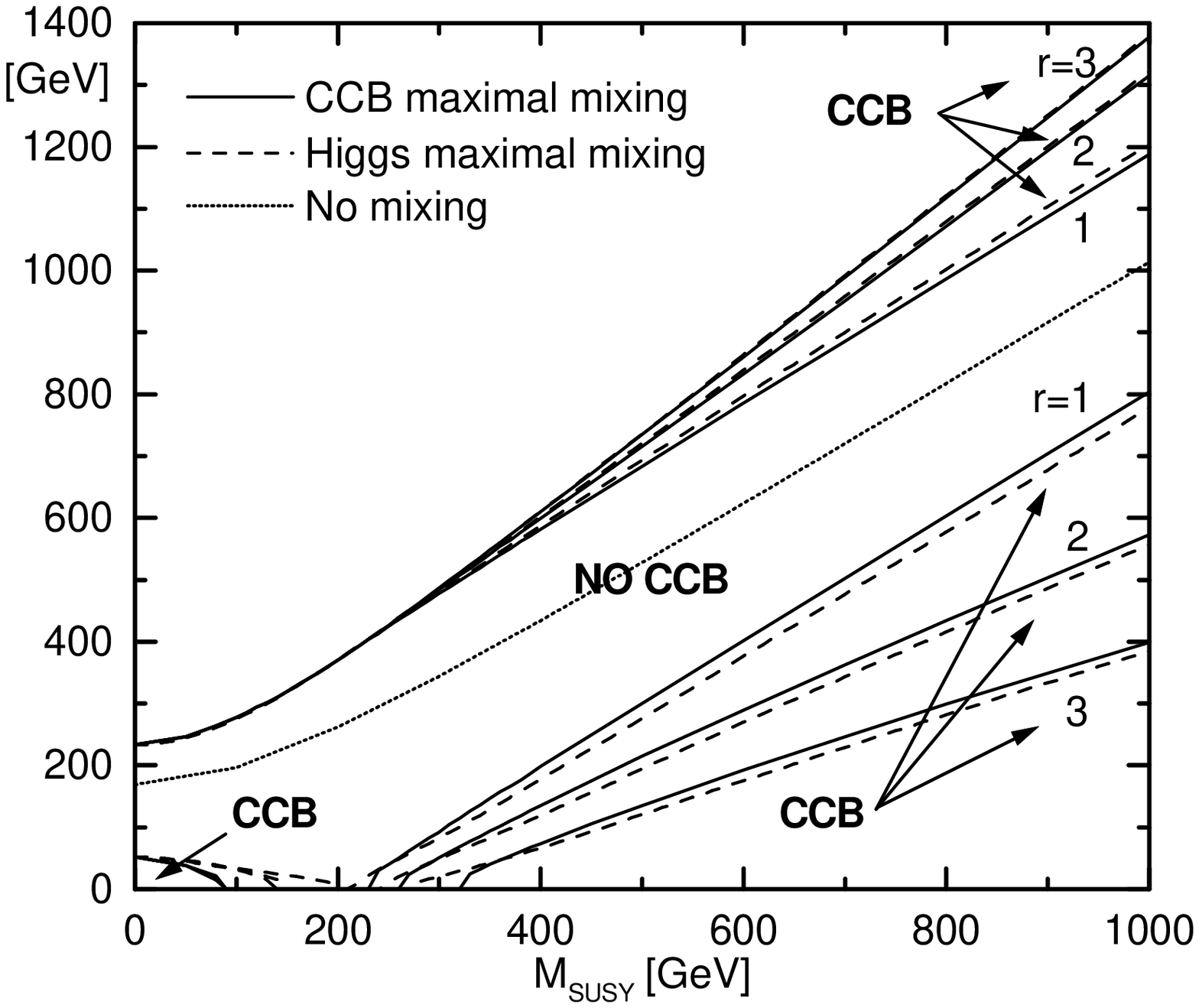,width=12.5cm}}
\end{center}
\vspace*{-1.9cm} \caption[Figure 4b]{The bounds on the stop masses
for the CCB maximal mixing and the Higgs maximal mixing, versus
$M_{SUSY}$, for $r=1,2,3$. Lines below [respectively, above] the
no mixing case give lower [respectively, upper] bounds  on the
lightest [respectively, heaviest] top squark mass.}
\end{figure}
The Higgs maximal mixing value $A_t^{H}$, which maximizes $m_h$ in
eq.(\ref{bmh}), can be obtained numerically as a function of
$M_{SUSY}$ and $r$. As is well-known, for large
$M_{SUSY}=m_{\tilde{t}_L}=m_{\tilde{t}_R} \gg m_t$, $A_t^H$ takes
the simple expression: $A_t^H= \sqrt{6} m_{\tilde{t}}$
\cite{higgs1,higgs2} [this value is actually used as a
normalization factor in Figs.1-2]. Indeed, this asymptotic
behaviour is observed in Fig.2 for $r=1$. In addition, Fig. 2
shows that $A_t^H$ is decreasing for an increasing splitting
between the soft squark masses \cite{higgs1}.\\
 For low $M_{SUSY}$,
the CCB maximal mixing $A_t^{CCB}$ and the Higgs maximal mixing
$A_t^H$ follow each other closely, showing that $m_h$ is maximal
for $A_t \simeq A_t^{CCB}$ [we have furthermore $A_t \simeq
A_t^{inst}$]. Notice that $A_t^H$ can be lower than $A_t^{CCB}$
for $M_{SUSY} \lsim m_t$, though just slightly and, moreover, in
an unphysical region where the lightest stop mass is vanishing
$m_{\tilde{t}_1} \simeq 0 \ GeV$ [see Fig.3] \footnote{Some
residual gauge contributions are actually neglected in the writing
of eq.(\ref{bmh}) in order to gain approximate independence for
$m_h$ with respect to the renormalization scale $Q$ \cite{higgs1}.
For low $M_{SUSY}$, such contributions may presumably restore the
hierarchy $A_t^H \ge A_t^{CCB}$.}. For larger values of
$M_{SUSY}$, the CCB maximal mixing clearly rules out the Higgs
maximal mixing. For $M_{SUSY}=1 \ TeV$, $A_t^{CCB}$ is about $10
\%$ below $A_t^H$, for $r=1,2,3$. Thus, the large exclusion
already observed in ref.\cite{ccb1} for equal soft squark masses
is also found in the presence of a large mass splitting.\\
\begin{figure}[htb]
\vspace*{-1.4cm}
\begin{center}
\mbox{ \psfig{figure=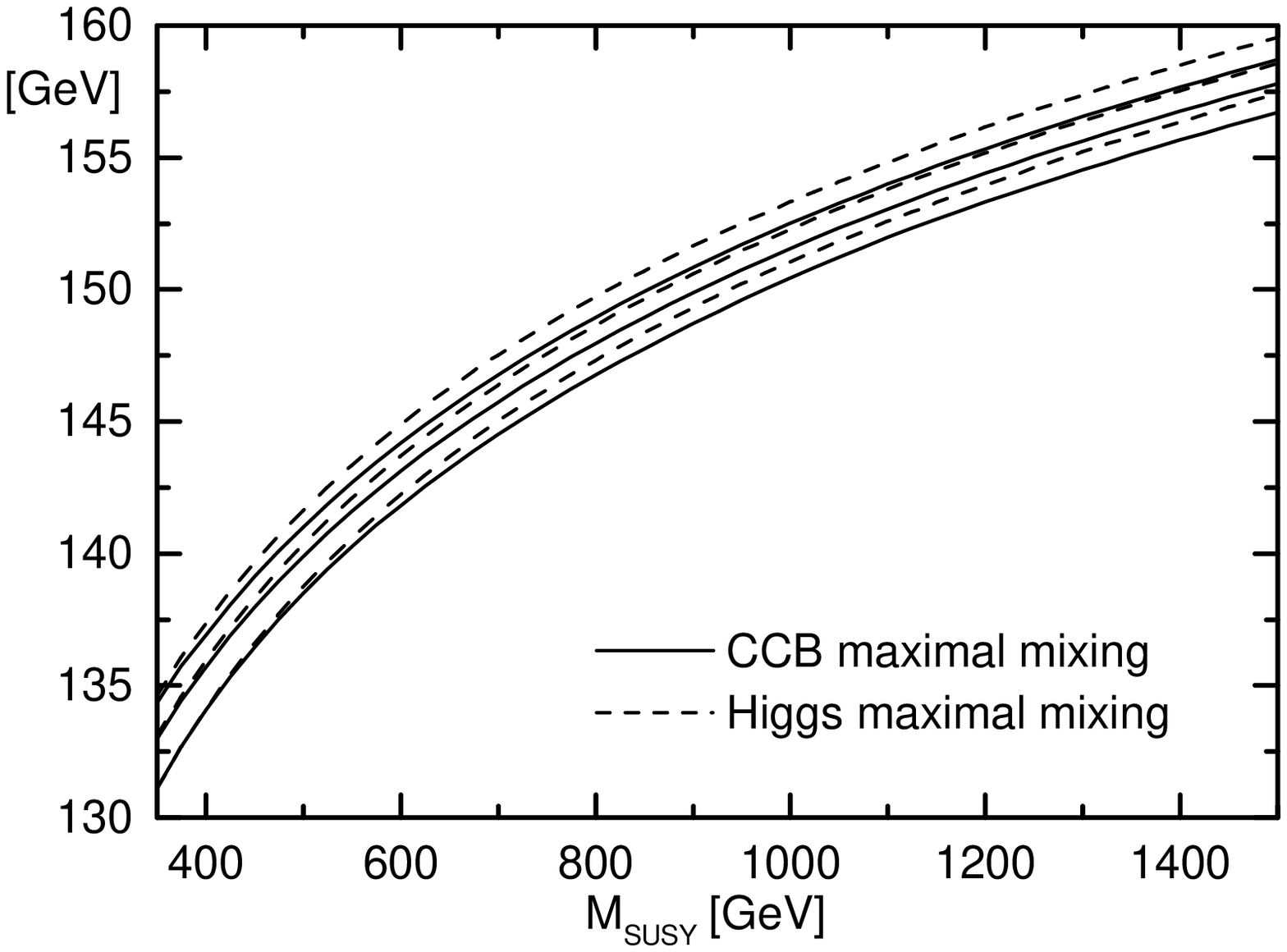,width=12.5cm}}
\end{center}
\vspace*{-1.cm} \caption[Figure 4b]{The one-loop upper bounds on
$m_h$ for the CCB maximal mixing and the Higgs maximal mixing,
versus $M_{SUSY}$. The higher, intermediate and lower lines
correspond, respectively, to $r=1,2$ and $3$.}
\end{figure}

In Fig.3, we compare the bounds on the top squark mass spectrum
for the CCB and the Higgs maximal mixing values. In both cases, we
display below [respectively, above] the no-mixing line, i.e.
$m_{\tilde{t}_{1},\tilde{t}_2}^2= M_{SUSY}^2+ m_t^2 -
m_{Z^0}^2/4$, the corresponding lower [respectively, upper] bounds
on the lightest stop mass $m_{\tilde{t}_1}$ [respectively, the
heaviest stop mass $m_{\tilde{t}_2}$]. For large $M_{SUSY}$, the
allowed range for the top squark mass spectrum is enlarged for an
increasing splitting between the soft squark masses, despite the
decrease of the CCB and the Higgs maximal mixing values for
increasing $r$ [see Fig.2].  Obviously, the CCB bounds on the top
squark mass spectrum are more restrictive than the Higgs maximal
mixing ones. This effect is however not very large, although not
negligible for $r=1$. For instance, for $M_{SUSY}= 1 \ TeV$, we
have respectively $\Delta m_{\tilde{t}_{1}} \simeq (25.5, 16.5,
14.5) \ GeV$ and $\Delta m_{\tilde{t}_{2}} \simeq (17, 7 , 4) \
GeV$, for $r=1,2,3$.\\
 Fig.3 exhibits another interesting feature.
Taking conservatively $m_{\tilde{t}_{1}} \gsim 100 \ GeV$ as an
experimental limit on the lightest stop mass \cite{stop1}, we find
that a stop mixing value as large as the CCB maximal mixing is
excluded in a large part of the parameter space, i.e. $M_{SUSY}
\lsim (310,$ $ 360,440) \ GeV$ for $r=1,2,3$. In the respective
domains, the EW vacuum is not threatened by the CCB vacuum in the
plane $(H_2,\tilde{t}_L,\tilde{t}_R)$, and is automatically
stable. This result illustrates how a precise study of CCB
conditions can produce refined statements concerning metastability
of the EW vacuum\footnote{In the metastability domain, we further
remark that the analytic expressions for the VEV's of the CCB
vacuum presented in the last section should also be very useful in
precisely evaluating the CCB metastability condition on $A_t$
\cite{ccbm}, for large $\tan \beta$.} \cite{ccbm}.\\

In Fig.4, we finally compare the one-loop upper bound on the
CP-even Higgs boson $m_h$ in the top-stop approximation,
eq.(\ref{bmh}), for the CCB and the Higgs maximal mixing values.
In both cases, this bound is decreasing with increasing $r$, but
the effect is rather small, at most $\sim 2-3 \ GeV$. The mass
discrepancy between the CCB and the Higgs maximal mixing cases is
negligible for $M_{SUSY} \sim 500 \ GeV$. It is slowly increasing
with $M_{SUSY}$, but is still small for $M_{SUSY} = 1500 \ GeV$,
where it is $ \sim 1 \ GeV$ for $r=1,2,3$. Let us note that in the
large $\tan \beta$ regime investigated here, $m_h$ may also
receive at one-loop level important additional contributions,
coming in particular from the bottom/sbottom sector
\cite{higgs1,higgs2}. Such contributions would modify the
numerical upper bound on $m_h$, but not the discrepancy between
the CCB and the Higgs maximal mixing values presented here, which
depends essentially on the top-stop contribution.\\

Thus, the numerical benefit of taking the CCB maximal mixing
rather than the exact one-loop Higgs maximal mixing to constrain
the top squark mass spectrum and the one-loop upper bound on
$m_h$, is not very large (although not always negligible).
However, on theoretical ground, this statement must be completed
by stressing that the requirement of avoiding a dangerous CCB
vacuum provides a strong and independent physical motivation to
consider top squark mixing terms smaller than the Higgs maximal
mixing. The latter in contrast represents a benchmark mixing,
useful essentially in keeping track of the value that maximizes
the lightest CP-even Higgs boson mass. Moreover, outside this
context, the CCB maximal mixing can have more drastic
phenomenological implications. For instance, at the tree-level, it
was shown that the cross section of production of the lightest
CP-even Higgs boson $h$ in association with a lightest stop pair
is strongly enhanced for a large stop mixing term $|\tilde{A}_t|
\sim \sqrt{6} m_{\tilde{t}}$ and can even exceed the production
cross section in association with a top quark pair at the CERN
Large Hadron Collider \cite{pheno2}. However, this interesting
window for the discovery of supersymmetric particles opens for a
lightest stop mass light enough at $m_{\tilde{t}_1} \sim m_t$,
therefore in a region of the parameter space where the optimal CCB
conditions are very restrictive. Clearly, Figs.2-3 show that the
two requirements: a light stop mass $m_{\tilde{t}_1} \sim m_t$ and
a large stop mixing $|\tilde{A}_t| \sim \sqrt{6} m_{\tilde{t}}$
are in conflict [we note that CCB occurs in the plane $(H_1,
H_2,\tilde{t}_L, \tilde{t}_R)$ for $|\tilde{A_t}| \ge
A_t^{CCB}|_{\tan \beta=+\infty}$, eq.(\ref{at2})]. Hence, we
expect a dramatic reduction of the cross section of such a process
in the CCB allowed region of the parameter space\footnote{It
remains to be determined if in some regions of the parameter
space, this process is still favored compared to the production
rate in association with a top quark pair. This will be the
subject of future investigations.}.\\
 This example illustrates the phenomenological usefulness
of the CCB maximal mixing $A_t^{CCB}$ for $\tan \beta=+ \infty$,
considered in this paper. As noted before, this benchmark mixing
can also be used to avoid metastability of the EW vacuum in
model-dependent scenarios, which unavoidably occurs if
$|\tilde{A}_t| \ge A_t^{CCB}|_{\tan \beta=+ \infty}$ at the SUSY
scale. In these contexts, the simple approximation
$A_t^{CCB}|_{app.}$, eq.(\ref{atap}), should be of particular
interest. Moreover, it is definitely more reliable than the
traditional CCB bound in the D-flat direction $A_t^{D}$,
eq.(\ref{condfrere}), often considered as a first guess of the
impact of CCB conditions, but which largely underestimates the
restrictive power of the latter.\\

 Finally, we remark that two-loop contributions provide
important contributions to $m_h$ and induce a displacement of the
Higgs maximal mixing, which may become more restrictive than the
CCB maximal mixing $A_t^{CCB}$: for $m_{A^0}, M_{SUSY} \gg
m_{Z^0}$, $\tan \beta=+ \infty$ and $r=1$, $A_t^H|_{2-loop} \simeq
2 \ m_{\tilde{t}}$ \cite{higgs2}, whereas $A_t^{CCB} \simeq 2.17 \
m_{\tilde{t}}$. However, at two-loop level, a precise
investigation of the effect of CCB conditions on the Higgs boson
mass $m_h$ requires the evaluation of the one-loop CCB bound
$A_t^{CCB}|_{1-loop}$. This value incorporates in particular
contributions which escape our tree-level improved CCB bound
$A_t^{CCB}$, evaluated at the SUSY scale. Therefore, the previous
comparison seems somewhat misleading. However, it raises the
important question of the hierarchy between $A_t^{CCB}|_{1-loop}$
and $A_t^H|_{2-loop}$. For low $M_{SUSY} \lsim m_t$, we remark
that the relation $A_t^{CCB} \simeq A_t^{inst} \simeq A_t^H$ [see
Figs.1-2] should persist at the next loop level. For large $\tan
\beta$, it is due essentially to the presence of an interference
regime where the CCB vacuum and the EW vacuum overlap. At the
one-loop level, the EW vacuum is still driven in the plane
$(H_2,\tilde{t}_L, \tilde{t}_R)$ for large $\tan \beta$, and such
a regime should therefore also be found. For $M_{SUSY} \gg m_t$,
things are not so clear. However, we may reasonably expect that
one-loop corrections will also lower the CCB maximal mixing, as
occurs for the Higgs maximal mixing, implying presumably the
hierarchy $A_t^{CCB}|_{1-loop} \le A_t^H|_{2-loop}$. This
expectation can be checked only by a complete one-loop
investigation of the CCB conditions in the plane
$(H_2,\tilde{t}_L, \tilde{t}_R)$, for large $\tan \beta$, which we
plan to do in the future.

\section{Conclusion}

In this paper, we investigated the optimal CCB condition on $A_t$
in the plane $(H_2,\tilde{t}_L, \tilde{t}_R)$, taking into account
the possibility of a large mass splitting between the soft squark
masses $m_{\tilde{t}_L},m_{\tilde{t}_R}$, as occurs in interesting
models \cite{ans,anom}. We essentially focused on the asymptotic
regime $\tan \beta=+ \infty$, motivated by the nice features of
the CCB bound in this regime. In particular, a complete
investigation of CCB conditions in the extended plane
$(H_1,H_2,\tilde{t}_L, \tilde{t}_R)$, which will be presented
elsewhere \cite{ccb3}, shows that the stop mixing term
$\tilde{A}_t$, in absolute value, should not exceed this benchmark
value, otherwise CCB unavoidably occurs. This CCB bound should
therefore be useful for phenomenological applications. For this
reason, we presented an accurate analytic approximation for it,
which fits the exact result within less than $1 \ \%$ in the
interesting phenomenological region where $m_{\tilde{t}_1} \gsim
100 \ GeV$.\\
 For $M_{SUSY} \gg m_t$, we showed that the one-loop Higgs maximal
mixing is ruled out by more than $10 \ \%$, whatever the splitting
between the soft squark masses $m_{\tilde{t}_L},m_{\tilde{t}_R}$
is. Compared to the Higgs maximal mixing, the effect of the CCB
maximal mixing on the top squark mass spectrum and on the one-loop
upper bound on $m_h$ is not very large, though not always
negligible. We pointed out however that larger effects can be
expected in Higgs phenomenology, which can be more sensitive to
such an exclusion. Further investigations are in progress in this
direction.\\

\noindent{\bf Acknowledgement}\\

\noindent This work was supported by a Marie Curie Fellowship,
under contract No HPMF-CT-1999-00363. I would like to thank G.J.
Gounaris and P.I. Porfyriadis for discussions.  Special thanks
also go to G. Moultaka for useful comments and for reading the
manuscript.

\end{document}